\documentclass[conference]{IEEEtran}
\IEEEoverridecommandlockouts
\usepackage{cite}
\usepackage{amsmath,amssymb,amsfonts}
\usepackage{graphicx}
\usepackage[export]{adjustbox}
\usepackage{textcomp}
\usepackage{xcolor}
\usepackage{float}
\usepackage{subcaption}  % 核心包
\usepackage[noend]{algpseudocode}
\usepackage{algorithmicx,algorithm}
\usepackage{caption}
\usepackage{multirow}
\usepackage{makecell}
\def\BibTeX{{\rm B\kern-.05em{\sc i\kern-.025em b}\kern-.08em
    T\kern-.1667em\lower.7ex\hbox{E}\kern-.125emX}}
\begin{document}

\renewcommand{\arraystretch}{1.15}

\title{WiFi-based Multi-user Activity Recognition via User-Conditioned Spatial Attention}

\author{\IEEEauthorblockN{Chenhan Yuan, Ruijing Liu, Cunhua Pan\thanks{Chenhan Yuan, Ruijing Liu, Cunhua Pan are with the National Mobile Communications Research Laboratory, Southeast University, Nanjing 210096, China (e-mail: yuanch@seu.edu.cn; 230258158@seu.edu.cn; cpan@seu.edu.cn).}}
%\IEEEauthorblockA{\textit{National Mobile Communications Research Laboratory} \\
%\textit{Southeast University}
%}
%\textsuperscript{nd}
}

\maketitle

\begin{abstract}
WiFi-based human activity recognition has achieved high accuracy under the single-user scene. Recognizing activities performed by multiple concurrent users remains challenging, because their body-reflected propagation paths superimpose in the channel state information (CSI) measurement.
Existing multi-user methods either decompose the signal before classification, suffering from error propagation, or predict activities jointly without reliably associating each prediction with the correct user.
Moreover, attention mechanisms that have proven effective for single-user sensing compute a single user-agnostic attention map, blending patterns from different users and providing no mechanism to associate features with specific user slots.
This paper proposes a user-conditioned spatial attention (UCSA) module that generates per-user spatial attention maps by conditioning multi-scale spatial attention on learnable user-ID embeddings via feature-wise linear modulation (FiLM).
UCSA creates a deep coupling between spatial feature enhancement and user differentiation: the attention module itself becomes user-aware, producing distinct attention maps for each user slot rather than relying on a late-fusion embedding addition.
Combined with a shareable multi-semantic spatial attention (SMSA) module for multi-scale spatial enhancement, the proposed framework addresses both feature quality and user differentiation in a unified architecture.
Experiments on the WiMANS benchmark across three indoor environments and up to five concurrent users show that the proposed method outperforms nine baselines in all settings, achieving an average accuracy of over 93\% across three environments even with five users.
Ablation studies confirm that both SMSA and UCSA contribute substantially to recognition accuracy, with UCSA providing the critical link between shared features and per-user predictions.
\end{abstract}

\begin{IEEEkeywords}
WiFi sensing, multi-user activity recognition, deep learning, user-conditioned attention, user-ID embedding, channel state information
\end{IEEEkeywords}

\section{Introduction}

WiFi-based human activity recognition (HAR) has attracted growing interest because it operates through walls, requires no wearable sensors, and leverages ubiquitous wireless infrastructure~\cite{yousefi2017survey,ma2020wifi,wu2017device}.
The underlying principle is that body movement affects the channel state information (CSI) of indoor WiFi links, and these effects carry motion signatures that can be classified by machine learning models.
Deep learning has substantially advanced single-user HAR: convolutional and recurrent architectures now achieve high accuracy on gesture and activity benchmarks~\cite{ma2018signfi,li2021twostream,yao2019stfnets}, while cross-domain methods have improved environment robustness~\cite{zhang2021widar3,yin2024fewsense}.

Despite this progress, most existing methods assume that only one user occupies the sensing region.
In practical indoor environments, multiple users are frequently present, and their body-reflected propagation paths superimpose in the CSI measurement.
This inter-user interference creates two coupled challenges: the shared CSI signal blends motion signatures from different users, and the model must assign each recognized activity to the correct individual without prior location information.
Existing multi-user approaches either decompose the signal before classification, which suffers from error propagation, or train end-to-end models that predict activities jointly but cannot reliably associate predictions with specific users~\cite{duan2023wisdom,venkatnarayan2021multiuser,abuhoureyah2024multi}.

Attention mechanisms have proven effective for adaptive feature recalibration in WiFi sensing~\cite{gu2022wigrunt,liu2025wifi}, yet existing studies have been developed and validated exclusively in single-user settings.
When multiple users are present, spatial attention maps inevitably blend patterns from different users, producing a single user-agnostic saliency map that cannot distinguish which spatial regions correspond to which user.
Therefore, attention alone cannot differentiate user identities: it enhances feature quality but provides no mechanism for associating features with specific user slots.
Consequently, the key challenge of applying attention in multi-user scenarios is to make the attention module itself user-aware and to inject user identity into the feature extraction stage rather than relying on late fusion.

\begin{figure}[!t]
    \centering
    \includegraphics[width=\columnwidth]{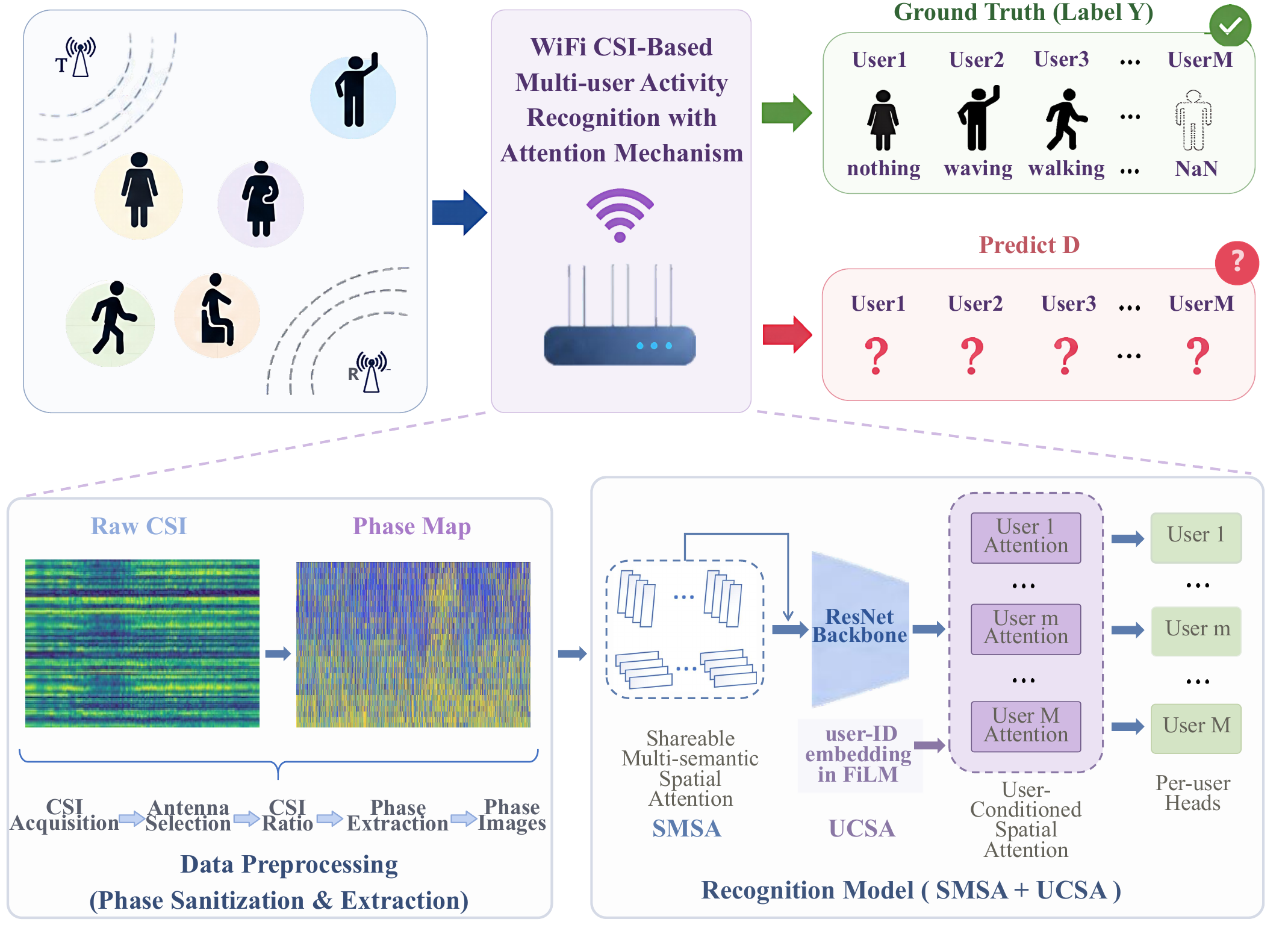}
    \caption{The overall framework of WiFi-based multi-user activity recognition system.}
    \label{fig:overallframework}
\end{figure}

As illustrated in Fig.~\ref{fig:overallframework}, we propose a unified framework for multi-user activity recognition scenario that integrates a user-conditioned spatial attention (UCSA) module and a shareable multi-semantic spatial attention (SMSA) module. 
The UCSA module creates a deep coupling between spatial feature enhancement and user differentiation. It generates per-user spatial attention maps by conditioning multi-scale spatial attention on learnable user-ID embeddings via FiLM-style modulation~\cite{perez2018film}: for each user slot, the user-ID embedding produces scale and shift parameters that modulate a shared base spatial attention map, yielding a distinct attention pattern that highlights the spatial regions most relevant to that user. 
This design ensures that the attention module itself becomes user-aware, rather than treating user differentiation as a post-attention add-on. 
The SMSA module complements UCSA by providing shared multi-scale spatial feature enhancement from CSI phase-ratio images. Together, they enable the proposed framework to address both feature quality and user differentiation in a unified architecture.
We evaluate the proposed method on the WiMANS benchmark~\cite{huang2024wimans} across three indoor environments and six user-count configurations. 
Experimental results show that the proposed method outperforms nine baselines in all settings, achieving an average accuracy of over 93\% across three environments even with five users. Ablation studies confirm that both SMSA and UCSA contribute substantially, with UCSA providing the critical link between shared spatial features and per-user predictions.

The main contributions of this work are as follows:
\begin{itemize}
\item We propose a UCSA module that generates per-user spatial attention maps by conditioning multi-scale spatial attention on learnable user-ID embeddings via feature-wise linear modulation (FiLM), creating a deep coupling between spatial feature enhancement and user differentiation.
\item We design an end-to-end framework for WiFi-based multi-user activity recognition that combines shared spatial enhancement with user-aware UCSA module, enabling data-driven user slot differentiation without ground-truth location information.
\item We evaluate the proposed framework on the WiMANS dataset, and achieve an average accuracy of over 99\% in the single-user case and over 93\% even with five concurrent users, outperforming nine baseline methods across three indoor environments. 
\end{itemize}

\section{RELATED WORK}

\subsection{WiFi-based human activity recognition.}
WiFi-based HAR extracts human motion signatures from CSI variations caused by body-reflected propagation paths~\cite{yousefi2017survey,ma2020wifi,wu2017device}.
Early methods relied on model-driven features or handcrafted features tailored to specific gestures ~\cite{li2016wifinger,qian2017widar,qian2018widar2}, which offered interpretability but limited generalization across environments and activities.
The advent of deep learning dramatically improved recognition performance: CNN-based models captured spatial variation patterns from CSI spectrograms~\cite{ma2018signfi}, while RNN and CNN-RNN hybrid architectures added temporal modeling to recognize sequential activities ~\cite{li2021twostream,yao2019stfnets}.
Large-scale benchmarks such as SenseFi~\cite{yang2023sensefi} and cross-domain methods like Widar3.0~\cite{zhang2021widar3} and FewSense~\cite{yin2024fewsense} further advanced the field by enabling systematic evaluation and environment-robust recognition.
However, the above methods are predominantly designed under the single-user assumption; when multiple users occupy the sensing region simultaneously, their reflected signals superimpose in the CSI measurement, causing severe inter-user interference that single-user models cannot disentangle.

\subsection{Multi-user activity recognition.}
Recognizing activities performed by multiple concurrent users introduces the fundamental challenge of separating overlapping motion signatures from a shared CSI stream.
Existing approaches can be broadly divided into two categories.
The first category pursues \emph{signal decomposition}: WISDOM~\cite{duan2023wisdom} sorts CSI substreams by signal quality before per-user classification, MultiTrack~\cite{tan2019multitrack} reconstructs individual link profiles for tracking and activity recognition, and IMar~\cite{he2022imar} factorizes the CSI tensor into user-specific components.
These methods achieve reasonable separation but suffer from error propagation: inaccuracies in the decomposition stage directly degrade downstream classification.
The second category adopts \emph{joint learning}: MultiSense~\cite{venkatnarayan2021multiuser} and WiMAR~\cite{abuhoureyah2024multi} train end-to-end networks that simultaneously predict activities for all users.
While avoiding signal separation, these methods lack an explicit mechanism to associate each prediction with the correct user, making them sensitive to user permutations and unable to exploit user-specific patterns.
The WiMANS benchmark~\cite{huang2024wimans} has recently provided a standardized dataset for reproducible multi-user evaluation, yet current methods on this benchmark still rely on either error-prone decomposition or permutation-ambiguous joint prediction.

\subsection{Attention mechanisms for WiFi sensing.}

In WiFi sensing, WiGRUNT~\cite{gu2022wigrunt} introduced a Spatial-Channel Synergistic Attention (SCSA) module that applies multi-scale depthwise convolutions along horizontal and vertical directions to capture motion patterns at different spatial granularities.
More recently, Liu~\emph{et~al.}~\cite{liu2025wifi} combined a SMSA module with a self-attention-based channel attention mechanism for cross-domain gesture recognition, demonstrating that spatial-channel attention synergies significantly improve feature discrimination under domain shift.
However, both methods were developed and validated exclusively in single-user settings.
When multiple users are present, the inter-user signal interference dilutes spatial attention selectivity and confuses channel-wise importance estimation, as the attention maps inevitably blend patterns from different users.
Moreover, attention alone cannot differentiate user identities: it enhances feature quality but provides no mechanism for associating features with specific user slots.
A method that conditions spatial attention on user identity information, so that each user slot receives a distinct attention map, is therefore needed to enable reliable multi-user activity recognition.

\section{SYSTEM DESIGN}

This section describes the preprocessing pipeline for raw CSI data and presents the attention-based network architecture for multi-user activity recognition.
The CSI preprocessing module extracts phase ratio features from raw CSI measurements and converts them into time-series images.
The activity recognition module employs SMSA module for shared multi-scale spatial feature enhancement, followed by UCSA module that generates per-user spatial attention maps conditioned on learnable user-ID embeddings, and per-user classification heads for multi-user activity prediction.

\subsection{Data Preprocessing}

\subsubsection{WiFi Channel State Information}

We consider a WiFi OFDM system operating in the 5~GHz band over a $20\,\mathrm{MHz}$ channel with $N=30$ active subcarriers.
The CFR at the $k$-th subcarrier is decomposed into a static and a dynamic component:
\begin{equation}
	H(k,t) = H_{\mathrm{s}}(k) + H_{\mathrm{d}}(k,t),
\end{equation}
where $H_{\mathrm{s}}(k)$ is the time-invariant component from fixed scatterers, and $H_{\mathrm{d}}(k,t)$ captures time-varying reflections introduced by human activity.
The dynamic component is modeled as the superposition of $L$ dynamic paths~\cite{gu2022wigrunt}:
\begin{equation}
	 H_{\mathrm{d}}(k,t) = \sum_{l_d=1}^{L_d} a_{l_d}(k) \, e^{-j 2\pi \frac{d_{l_d}(t)}{\lambda_k}},
\end{equation}
where $a_{l_d}(k)$ is the complex amplitude of the $l_d$-th dynamic path, $d_{l_d}(t)$ denotes the time-varying path length change of the $l_d$-th dynamic path due to human motion, and $\lambda_k$ is the wavelength at the $k$-th subcarrier.
In multi-user scenarios, these dynamic paths arise from the reflections of all moving individuals, and the resulting phase variation inherently encodes the superimposed activity signatures of multiple users.

We process CSI phase rather than amplitude for activity recognition.
Phase is highly sensitive to path length changes: a displacement of $\Delta d = \lambda/4 \approx 1.5\,\mathrm{cm}$ at $5\,\mathrm{GHz}$ produces a full $\pi$ phase rotation, whereas the same displacement causes only a minor amplitude variation.
In multi-user scenarios, different individuals may produce similar amplitude attenuation patterns due to coherent multipath superposition, whereas phase captures fine-grained differences in spatial position, motion velocity, and limb swing amplitude, providing more discriminative features for separating concurrent users.

\subsubsection{Phase Sanitization via CSI Ratio}

Our WiFi transceiver employs a $3 \times 3$ multiple-input multiple-output(MIMO) configuration.
For each transmit antenna, the raw CSI at subcarrier $k$ and time instant $t$ from receive antenna $r$ can be expressed as
\begin{equation}
	H_{r,k}(t) = |H_{r,k}(t)| \cdot e^{j\left(\psi_{r,k}(t) + \phi_{\mathrm{off}}(t)\right)},
\end{equation}
where $\psi_{r,k}(t)$ is the true propagation phase encompassing both static reflections and the aggregate contribution of all dynamic reflections from moving users, and $\phi_{\mathrm{off}}(t)$ is a random phase offset caused by carrier frequency offset (CFO), sampling frequency offset (SFO), and hardware imperfections.
Since all receive antennas on the same NIC share the same RF oscillator, $\phi_{\mathrm{off}}(t)$ is identical across antennas and constitutes a common-mode disturbance.

To eliminate this common-mode offset, we adopt the CSI-ratio approach~\cite{gu2022wigrunt}.
We first select the best and worst receive antennas using the Quality Factor Metric (QFM), which evaluates antenna signal quality based on the ratio of mean amplitude to amplitude variance across subcarriers, and then compute the complex ratio between the best-quality antenna and each of the two worst-quality antennas:
\begin{equation}
	R_{i,k}(t) = \frac{H_{r_{\mathrm{best}},k}(t)}{H_{r_{\mathrm{worst}_i},k}(t)},
\end{equation}
where $i \in \{1,2\}$ indexes the two worst antennas.
Substituting the expression for $H_{r,k}(t)$, the common offset cancels:
\begin{equation}
	R_{i,k}(t) = \frac{|H_{r_{\mathrm{best}},k}(t)|}{|H_{r_{\mathrm{worst}_i},k}(t)|} \cdot e^{j\left(\psi_{r_{\mathrm{best}},k}(t) - \psi_{r_{\mathrm{worst}_i},k}(t)\right)}.
\end{equation}
The differential phase is then extracted as
\begin{equation}
	\theta_{i,k}(t) = \angle\left(R_{i,k}(t)\right) = \psi_{r_{\mathrm{best}},k}(t) - \psi_{r_{\mathrm{worst}_i},k}(t).
\end{equation}

Since the static component is time-invariant, its contribution to the phase difference is a constant that does not affect the time-varying activity signature.
In multi-user scenarios, the dynamic reflections from all moving individuals coherently combine at the receiver, and $\psi_{r,k}(t)$ inherently encodes the composite effect of all users' activities.
Consequently, $\theta_{i,k}(t)$ implicitly captures the combined multi-user activity information.
Although $\theta_{i,k}(t)$ cannot be decomposed into a simple linear superposition of individual user contributions, this composite representation serves as an effective input for the subsequent attention-based network, which learns to disentangle multi-user activity patterns in a data-driven manner.

For each transmit antenna, the best receive antenna is selected as reference and paired with the second-best and worst, yielding two phase-ratio streams. Three transmit antennas thus give 3 × 2 = 6 streams per CSI snapshot.
Each stream $\theta_{i,k}(t)$ for $k = 1, \ldots, 30$ and $t = 1, \ldots, T$ forms a $30 \times T$ matrix, which we convert to a grayscale image via linear intensity mapping, where the vertical axis corresponds to subcarrier index and the horizontal axis to time.
The six images provide complementary spatial perspectives of the same multi-user activity event, enhancing robustness against antenna-specific noise, and are resized to $224 \times 224$ pixels to match the input dimensions of the pretrained CNN backbone.

\subsection{Activity Recognition Model}

\subsubsection{Overall Framework}

\begin{figure*}[!t]
    \centering
    \includegraphics[width=\textwidth]{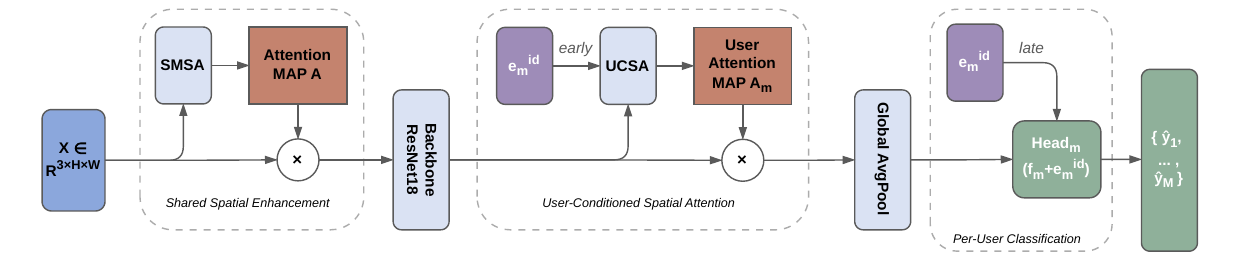}
    \captionsetup{justification=raggedright,singlelinecheck=false}
    \caption{The architecture of Attention-based Network.}
    \label{fig:model_architecture}
\end{figure*}

Fig.~\ref{fig:model_architecture} illustrates the architecture of the proposed multi-user activity recognition model.
The preprocessed CSI phase-ratio image $\mathbf{X} \in \mathbb{R}^{3 \times H \times W}$ is first enhanced by SMSA module that captures multi-scale spatial patterns along both axes.
The enhanced features are then fed into a pretrained ResNet18 backbone for hierarchical representation learning.
UCSA module takes the backbone output and a set of learnable user-ID embeddings $\mathbf{E}_{\text{id}} = [\mathbf{e}_1^{\text{id}}; \ldots; \mathbf{e}_M^{\text{id}}]$ to generate per-user spatial attention maps, where $M$ is the maximum number of concurrent users, equivalently, the maximum number of user slots.
Each user slot is then independently pooled and classified, with the user-ID embedding also injected at the classification head to create a dual interaction.
The key design principle is that user identity should enter the pipeline at the feature extraction stage rather than being deferred to the classifier, making the attention mechanism itself user-aware.

\begin{figure*}[!t]
    \centering
    \begin{subfigure}[b]{0.42\textwidth}
        \centering
        \includegraphics[trim={0 1cm 0 1cm}, clip, height=10.8cm]{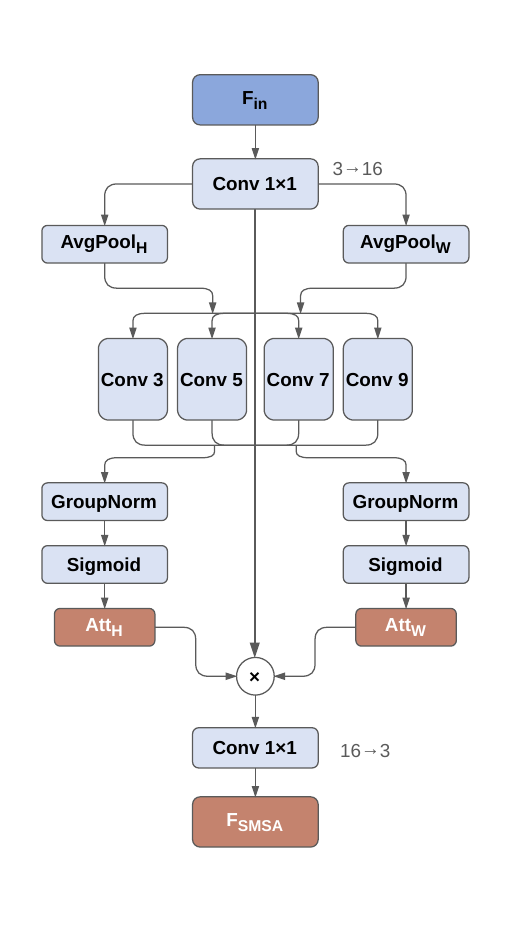}
        \caption{}
        \label{fig:smsa}
    \end{subfigure}
    \hfill
    \begin{subfigure}[b]{0.56\textwidth}
        \centering
        \includegraphics[trim={0 0.9cm 0 0.8cm}, clip, height=11cm] {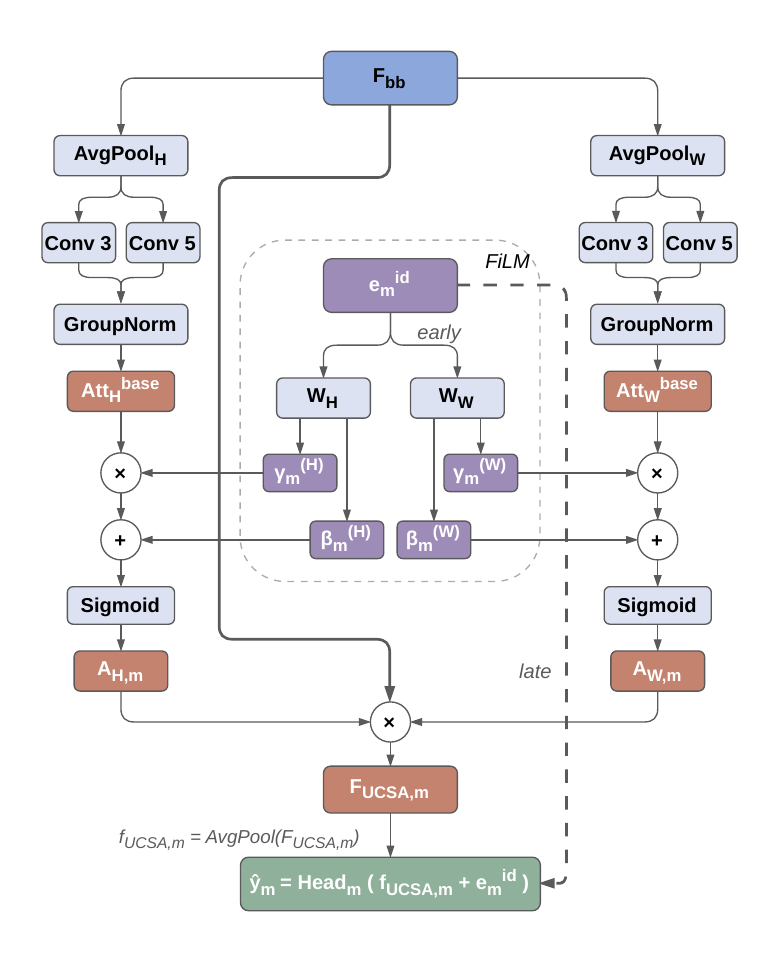} % trim 左 下 右 上
        \caption{}
        \label{fig:ucsa}
    \end{subfigure}
    \caption{(a) SMSA enhances shared spatial features via multi-scale convolution; (b) UCSA conditions the spatial attention on user identity to produce distinct per-user attention maps.}
    \label{fig:attention_modules}
\end{figure*}

\subsubsection{Multi-Scale Spatial Feature Enhancement}

CSI phase-ratio images exhibit strong spatial anisotropy: the vertical axis encodes subcarrier frequency, while the horizontal axis represents time. To exploit this structure, we adopt the SMSA module~\cite{liu2025wifi}, which extends the SCSA architecture~\cite{gu2022wigrunt} by performing bidirectional spatial attention along horizontal and vertical directions through parallel multi-scale branches. As illustrated in Fig.~\ref{fig:smsa}, the input feature map is first expanded from 3 to 16 channels via a \(1\times1\) convolutional layer, ensuring that the channels can be evenly divided into four groups. The expanded feature map is then averaged along each spatial axis to obtain 1-D descriptors, which are split into four channel groups. Each group is processed by a depthwise 1-D convolution with kernel size \(\{3, 5, 7, 9\}\), capturing motion patterns at progressively larger spatial scales---from fine-grained local gestures (\(k\!=\!3\)) to coarse-grained full-body movement (\(k\!=\!9\)). The convolved groups are concatenated, group-normalized, and passed through a sigmoid gate to produce the horizontal attention map \(\mathbf{A}_H \in \mathbb{R}^{16 \times H \times 1}\); a symmetric operation along the height dimension yields \(\mathbf{A}_W \in \mathbb{R}^{16 \times 1 \times W}\). The output is:
 \begin{equation}
 \mathbf{F}_{\text{SMSA}} = \mathbf{F}_{\text{in}} \odot \mathbf{A}_H \odot \mathbf{A}_W,
 \end{equation}
 where \(\odot\) denotes element-wise multiplication. The bidirectional decomposition aligns with the physical asymmetry of CSI images and reduces the attention complexity from \(O((H \!\times\! W)^2)\) to \(O(H^2 + W^2)\).

\subsubsection{UCSA Module}

In multi-user scenarios, the CSI phase-ratio image encodes the superimposed motion patterns of all active users: different users occupy different spatial regions and induce distinct time-frequency signatures.
SMSA, however, produces a single shared attention map that blends these patterns into a common saliency, providing no mechanism to associate spatial features with specific user slots.
A na\"ive remedy is to inject user identity only at the classification head (late fusion), yet by that point the spatial features have already been mixed across users, and the classifier cannot recover per-user information from a fused representation.
The central design choice of UCSA is therefore to make the attention mechanism itself user-aware: user identity should modulate spatial attention at the feature extraction stage, not merely inform the final classifier.

To this end, we define a learnable user-ID embedding matrix $\mathbf{E}_{\text{id}} \in \mathbb{R}^{M \times d}$, where $d$ matches the backbone output dimension ($d\!=\!512$).
Each row \(\mathbf{e}_m^{\text{id}}\) (\(m=1,\dots,M\)) serves as a slot-specific identifier that enables user differentiation in a purely data-driven manner, without requiring any ground-truth location input at inference time.

UCSA first computes a shared base spatial attention from the backbone output $\mathbf{F}_{\text{bb}} \in \mathbb{R}^{d \times H' \times W'}$.
Following the same H/W-axis decomposition as SMSA, the feature map is averaged along each spatial dimension, split into two channel groups, and processed by depthwise 1-D convolutions with kernel sizes $\{3, 5\}$.
After group normalization, this yields base attention maps $\mathbf{A}_H^{\text{base}} \in \mathbb{R}^{d \times H'}$ and $\mathbf{A}_W^{\text{base}} \in \mathbb{R}^{d \times W'}$ that capture multi-scale spatial saliency shared across all users.

For each user slot $m$, the user-ID embedding $\mathbf{e}_m^{\text{id}}$ is projected to produce FiLM-style~\cite{perez2018film} scale ($\boldsymbol{\gamma}$) and shift ($\boldsymbol{\beta}$) parameters, which modulate the base attention into a user-specific map, as illustrated in Fig.~\ref{fig:ucsa}:

\begin{equation}
\begin{aligned}
&\mathbf{A}_{H,m} = \sigma\!\big(\boldsymbol{\gamma}_m^{(H)} \odot \mathbf{A}_H^{\text{base}} + \boldsymbol{\beta}_m^{(H)}\big),\\
&\mathbf{A}_{W,m} = \sigma\!\big(\boldsymbol{\gamma}_m^{(W)} \odot \mathbf{A}_W^{\text{base}} + \boldsymbol{\beta}_m^{(W)}\big),
\end{aligned}
\end{equation}
where $[\boldsymbol{\gamma}_m^{(H)}, \boldsymbol{\beta}_m^{(H)}]= \mathbf{W}_H\, \mathbf{e}_m^{\text{id}} \in \mathbb{R}^{2d }$ and $[\boldsymbol{\gamma}_m^{(W)}, \boldsymbol{\beta}_m^{(W)}]= \mathbf{W}_W\, \mathbf{e}_m^{\text{id}} \in \mathbb{R}^{2d }$ , $\odot$ denotes element-wise multiplication, and $\sigma$ is the sigmoid function.
The scale parameter $\boldsymbol{\gamma}$ controls each channel's sensitivity to the base attention, while the shift parameter $\boldsymbol{\beta}$ offsets the activation threshold; together they enable each user slot to highlight a distinct spatial pattern from the same base saliency.
The per-user attended feature is then $\mathbf{F}_{\text{UCSA},m} = \mathbf{F}_{\text{bb}} \odot \mathbf{A}_{H,m} \odot \mathbf{A}_{W,m}$, and all $M$ feature maps are stacked into $\mathbf{F}_{\text{UCSA}} \in \mathbb{R}^{M \times d \times H' \times W'}$.

The user-slot embedding \(\mathbf{E}_{\text{id}}\) is shared across two injection points. In UCSA, \(\mathbf{e}_m^{\text{id}}\) is projected to FiLM parameters \((\gamma_m, \beta_m)\) that modulate the shared spatial attention. In the classification head, it is added to the pooled feature as a residual bias, i.e., \(\hat{y}_m = \text{Head}_m(\mathbf{f}_m + \mathbf{e}_m^{\text{id}})\), where \(\text{Head}_m\) is an independent linear classifier for slot \(m\). This dual-path injection allows the embedding to receive gradients from both the attention and classification branches, producing richer representations than a single-point injection. Unlike positional embeddings that encode geometric coordinates, the user-slot embeddings are learned end-to-end without any ground-truth location input. The shared base attention is computed once and modulated $M$ times through lightweight FiLM projections, adding minimal overhead compared to running $M$ independent attention branches.

\section{EXPERIMENTAL RESULTS}

\subsection{Dataset}

We evaluate the proposed method on the WiMANS benchmark~\cite{huang2024wimans}, the first publicly available dataset designed specifically for WiFi-based multi-user activity sensing.
CSI measurements were collected from three indoor environments---a classroom, an empty room, and a meeting room---using a $3\!\times\!3$ MIMO transceiver operating at $5\,$GHz.
We select the $5\,$GHz band because its shorter wavelength yields phase variations more sensitive to human-induced path changes, as analyzed in Section~III-A.
Five participants perform nine activities (\emph{nothing, walk, rotation, jump, wave, lie down, pick up, sit down, stand up}) at five predefined locations, with up to five users active simultaneously.
The dataset is randomly split into training and test sets at a $9{:}1$ ratio.
All reported results are the mean and standard deviation over five independent runs with different random seeds.

\subsection{Implementation Details}

The training objective minimizes the average cross-entropy loss over all valid (non-absent) user slots:
\begin{equation}
\mathcal{L} = \frac{1}{M_{\text{valid}}} \sum_{i=1}^{M} \mathbb{1}[p_i \neq -1] \cdot \text{CE}(\hat{y}_i, y_i),
\end{equation}
where $p_i$ is the presence indicator ($-1$ for absent users) and $M_{\text{valid}}$ counts active users in the batch. The per-slot averaging ensures that each valid user contributes equally to the gradient regardless of how many user slots are active.
The model is trained for $50$ epochs with a batch size of $16$ using the Adam optimizer (initial learning rate $10^{-3}$, step decay by $0.5$ every $5$ epochs).
Dropout with ratio $0.5$ is applied before the classification heads.
The SMSA module uses four depthwise convolution branches with kernel sizes $\{3,5,7,9\}$ and four attention heads.
The UCSA module uses two depthwise convolution branches with kernel sizes $\{3,5\}$ and FiLM-style conditioning from user-ID embeddings.
The user-ID embedding dimension is $d=512$, matching the ResNet18 backbone output dimension.
All input CSI images are resized to $224\!\times\!224$ pixels.

\subsection{Overall Performance}

\begin{table*}[!t]
\centering
\caption{Multi-seed recognition accuracy (\%) on the 5\,GHz WiMANS dataset across five random seeds.}
\label{tab:multi_seed}
\begin{tabular}{llccccc|c}
\hline
\textbf{Environment} & \textbf{User Count} & \textbf{Seed\,1} & \textbf{Seed\,2} & \textbf{Seed\,3} & \textbf{Seed\,4} & \textbf{Seed\,5} & \textbf{Mean$\pm$Std} \\
\hline
\multirow{6}{*}{Classroom}
 & 1 & 98.78 & 99.88 & 99.98 & 100.00 & 99.96 & 99.72$\pm$0.53 \\
 & 2 & 95.42 & 96.86 & 96.07 & 96.56 & 97.44 & 96.47$\pm$0.77 \\
 & 3 & 93.48 & 94.86 & 94.11 & 94.58 & 95.42 & 94.49$\pm$0.74 \\
 & 4 & 91.84 & 93.67 & 92.67 & 93.30 & 94.42 & 93.18$\pm$0.98 \\
 & 5 & 91.17 & 93.62 & 92.28 & 93.12 & 94.61 & 92.96$\pm$1.31 \\
 & All$^{\dagger}$ & 92.02&	94.21&	93.02&	93.76&	95.10 & 93.62$\pm$1.17 \\
\hline
\multirow{6}{*}{Empty room}
 & 1 & 98.89 & 99.53 & 99.18 & 99.40 & 99.79 & 99.36$\pm$0.34 \\
 & 2 & 95.04 & 96.20 & 95.57 & 95.96 & 96.67 & 95.89$\pm$0.62 \\
 & 3 & 91.84 & 93.73 & 92.70 & 93.34 & 94.49 & 93.22$\pm$1.01 \\
 & 4 & 90.77 & 92.92 & 91.75 & 92.48 & 93.79 & 92.34$\pm$1.15 \\
 & 5 & 89.31 & 92.47 & 90.75 & 91.82 & 93.75 & 92.10$\pm$1.69 \\
 & All &90.47&	93.34&	91.78&	92.75&	94.50& 92.57$\pm$1.53 \\
\hline
\multirow{6}{*}{Meeting room}
 & 1 & 99.88 & 99.77 & 98.83 & 99.62 & 99.99 & 99.62$\pm$0.46 \\
 & 2 & 94.91 & 95.83 & 95.33 & 95.64 & 96.20 & 95.58$\pm$0.49 \\
 & 3 & 94.33 & 94.80 & 94.54 & 94.70 & 94.99 & 94.67$\pm$0.25 \\
 & 4 & 92.80 & 94.39 & 93.52 & 94.06 & 95.03 & 93.96$\pm$0.85 \\
 & 5 & 91.19 & 94.08 & 92.51 & 93.48 & 95.24 & 93.30$\pm$1.54 \\
 & All &92.28&	94.38	&93.23&	93.94&	95.22& 93.81$\pm$1.12 \\
\hline
\multicolumn{8}{l}{\footnotesize $^{\dagger}$All: model trained on data from all user counts combined.} \\
\end{tabular}
\end{table*}

\begin{figure}[!t]
    \centering
    \includegraphics[width=\columnwidth]{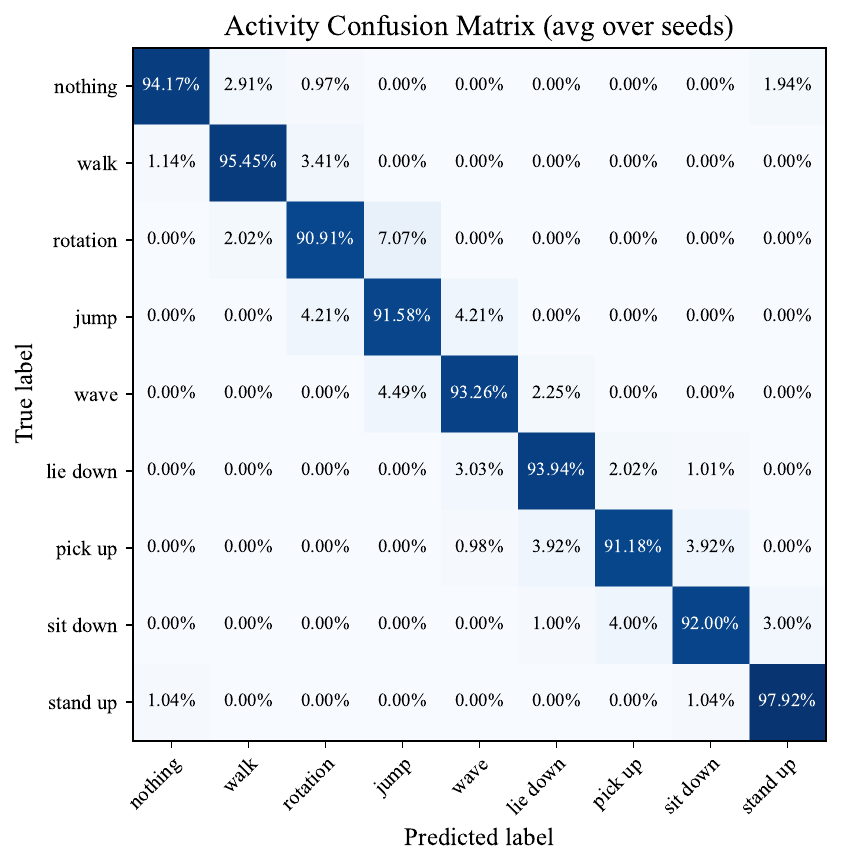}
    \caption{Activity confusion matrix (5-seed average) under the 5-user scenario.}
    \label{fig:confusion_activity}
\end{figure}
\begin{figure}[!t]
    \centering
    \includegraphics[scale=0.56]{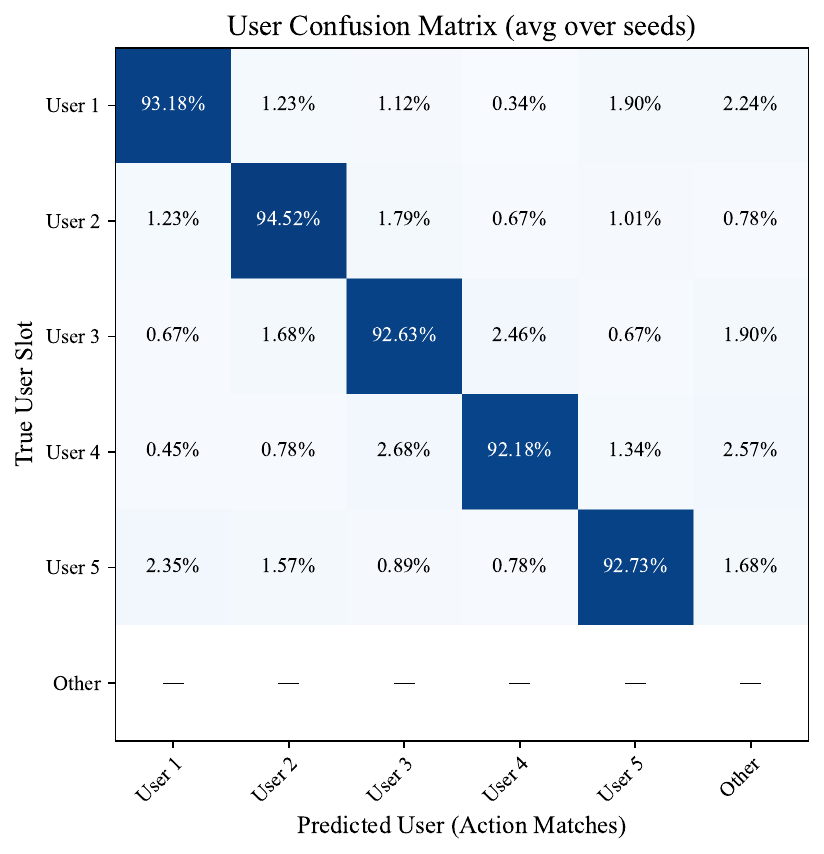}
    \caption{User-slot confusion matrix under the 5-user scenario. Rows: ground-truth user identity; columns: predicted user slot.}
    \label{fig:confusion_user}
\end{figure}

On the next page, Table~\ref{tab:multi_seed} reports the per-seed and mean accuracy for each environment under each number of concurrent users. The ``All`` setting denotes a single model trained on data from all user counts combined, rather than a separate model for each user count.
Across all $18$ conditions, the standard deviation remains below $1.7\%$, confirming that the model converges to consistent solutions regardless of random initialization.
In single-user scenarios the model exceeds $99\%$ accuracy in every environment, and even with five concurrent users it sustains above $91\%$.
The accuracy gap between single-user and five-user conditions is moderate ($7$--$8\%$), which is substantially smaller than the degradation observed in baseline methods (Table~\ref{tab:comparison}, on the next page).

Fig.~\ref{fig:confusion_activity} presents the activity-level confusion matrix averaged over five seeds under the 5-user scenario. Static and low-motion activities (\emph{nothing, walk, lie down, stand up}) are recognized with near-perfect accuracy, about 94\% to 97\%, as their CSI signatures differ markedly from other classes.
The primary confusion arises between \emph{rotation}/\emph{jump} and \emph{wave}/\emph{jump}: these pairs involve overlapping torso and limb trajectories that produce similar phase-ratio perturbations across subcarriers.

Fig.~\ref{fig:confusion_user} evaluates user-slot association accuracy, i.e., whether the model correctly assigns each user's activity prediction to the corresponding slot.
The diagonal-dominant structure indicates that the UCSA module, together with the user-ID embeddings, effectively differentiates user slots in a data-driven manner, which means differentiates user slots without relying on explicit location labels or hand-crafted rules.
The slight off-diagonal entries between P3 and P4 suggest that users at spatially proximate locations generate more similar CSI perturbations, consistent with the physics of overlapping multipath reflections.
Nevertheless, the model still assigns the majority of these cases correctly, indicating that the learned embeddings capture identity-discriminative patterns beyond pure spatial coincidence.

\begin{figure*}[tbp]
    \centering
    \includegraphics[width=\textwidth]{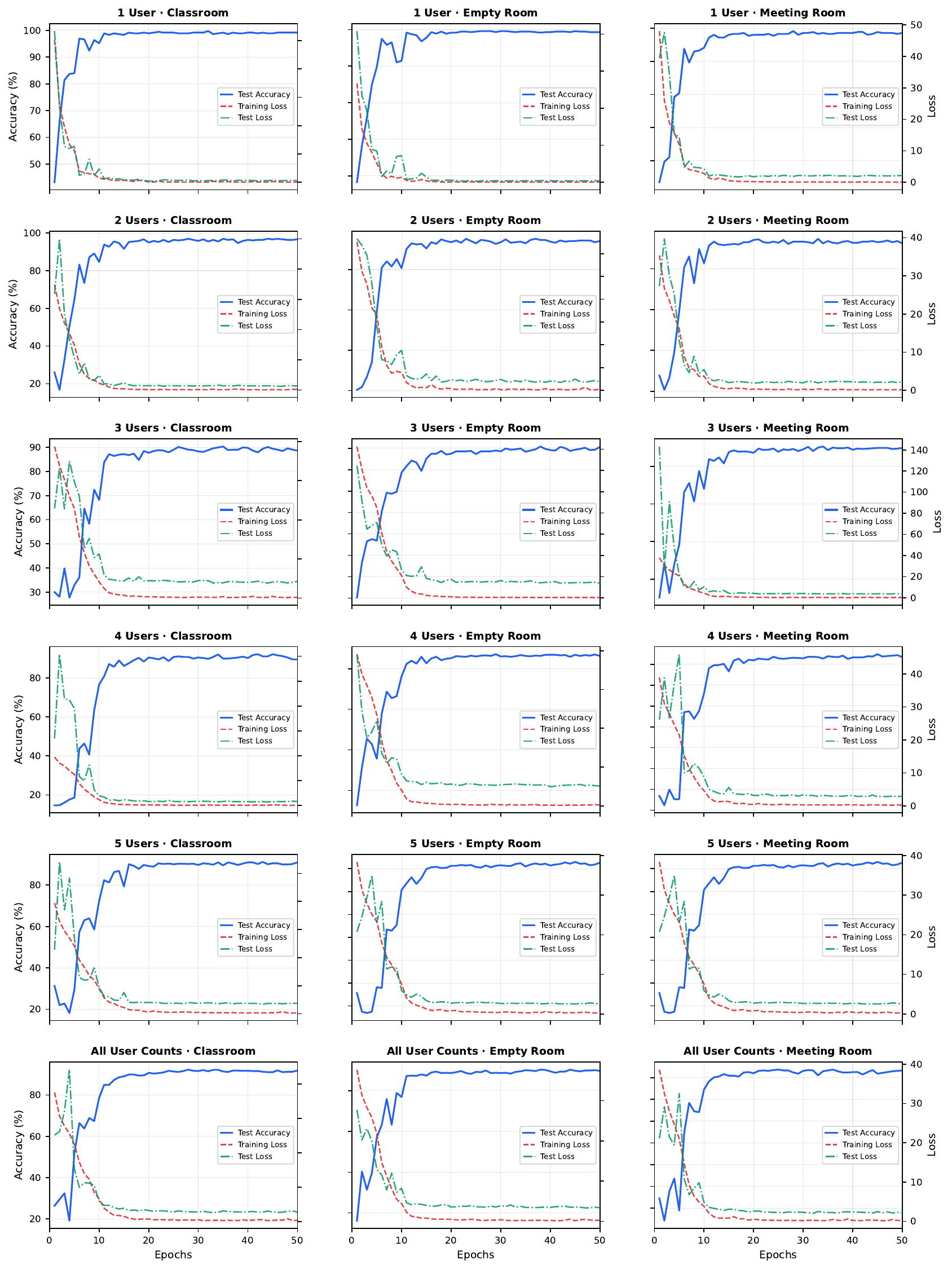}
    \captionsetup{justification=raggedright,singlelinecheck=false}
    \caption{Training loss, test loss, and test accuracy curves across three environments and six user configurations. }
    \label{fig:performance_grid}
\end{figure*}

We next examine the convergence behavior and overall accuracy. Fig.~\ref{fig:performance_grid} plots the training loss, test loss, and test accuracy curves across all three environments and six user configurations (1--5 users and all users combined).
Two trends are evident.
First, the model converges steadily within $30$--$40$ epochs across all settings, and the close alignment between training and test losses indicates no significant overfitting.
Second, recognition accuracy decreases gradually as the number of concurrent users increases: the additional body-reflected paths superimpose in the CSI measurement, creating stronger inter-user interference that makes individual activity signatures harder to isolate.
Notably, the all user-count configuration achieves accuracy comparable to the $4$--$5$ user conditions, demonstrating that a single model can handle mixed user-count scenarios without per-count specialization.

\subsection{Comparison with Baseline Methods}

\begin{table*}[!t]
\centering
\caption{Comparison of recognition accuracy (\%) across different methods. Mean$\pm$Std over 5 random seeds.}
\label{tab:comparison}
\resizebox{\textwidth}{!}{%
\begin{tabular}{llcccccccccc}
\hline
\textbf{Env.} & \textbf{User Count} & \textbf{ST-RF} & \textbf{MLP} & \textbf{LSTM} & \textbf{CNN-1D} & \textbf{CNN-2D} & \textbf{CLSTM} & \textbf{ABLSTM} & \textbf{THAT} & \textbf{ITA} & \textbf{Ours} \\
\hline
\multirow{6}{*}{Classroom}
 & 1 & 77.7$\pm$0.05 & 73.4$\pm$0.20 & 75.3$\pm$0.29 & 76.2$\pm$0.13 & 73.8$\pm$0.28 & 79.1$\pm$0.34 & 76.4$\pm$0.15 & 87.1$\pm$0.22 & 96.8$\pm$0.64 & \textbf{99.72$\pm$0.53} \\
 & 2 & 72.0$\pm$0.11 & 68.1$\pm$0.20 & 70.5$\pm$0.30 & 69.0$\pm$0.18 & 68.1$\pm$0.28 & 73.1$\pm$0.32 & 69.1$\pm$0.24 & 82.0$\pm$0.40 & 93.6$\pm$0.78 & \textbf{96.47$\pm$0.77} \\
 & 3 & 67.2$\pm$0.07 & 62.7$\pm$0.16 & 65.7$\pm$0.28 & 63.9$\pm$0.18 & 62.0$\pm$0.27 & 65.7$\pm$0.33 & 65.9$\pm$0.17 & 73.4$\pm$0.27 & 91.5$\pm$0.71 & \textbf{94.49$\pm$0.74} \\
 & 4 & 64.5$\pm$0.12 & 60.1$\pm$0.18 & 62.1$\pm$0.21 & 61.5$\pm$0.22 & 60.5$\pm$0.28 & 64.5$\pm$0.24 & 62.3$\pm$0.15 & 67.8$\pm$0.30 & 90.3$\pm$0.78 & \textbf{93.18$\pm$0.98} \\
 & 5 & 55.9$\pm$0.14 & 57.6$\pm$0.17 & 59.8$\pm$0.30 & 59.5$\pm$0.24 & 58.9$\pm$0.33 & 62.9$\pm$0.49 & 60.4$\pm$0.23 & 61.1$\pm$0.33 & 89.0$\pm$0.80 & \textbf{92.96$\pm$1.31} \\
 & All$^{\dagger}$ & 57.3$\pm$0.08 & 58.6$\pm$0.14 & 60.6$\pm$0.28 & 60.6$\pm$0.22 & 59.8$\pm$0.30 & 64.2$\pm$0.55 & 61.4$\pm$0.22 & 61.8$\pm$0.29 & 90.1$\pm$0.80 & \textbf{93.62$\pm$1.17} \\
\hline
\multirow{6}{*}{\makecell{Empty\\room}}
 & 1 & 80.8$\pm$0.08 & 74.0$\pm$0.18 & 75.1$\pm$0.27 & 75.7$\pm$0.28 & 77.1$\pm$0.18 & 76.2$\pm$0.40 & 78.7$\pm$0.24 & 86.9$\pm$0.21 & 97.5$\pm$0.80 & \textbf{99.36$\pm$0.34} \\
 & 2 & 75.0$\pm$0.11 & 71.6$\pm$0.22 & 70.3$\pm$0.19 & 69.2$\pm$0.19 & 70.8$\pm$0.26 & 75.8$\pm$0.34 & 69.0$\pm$0.25 & 83.1$\pm$0.32 & 94.1$\pm$0.85 & \textbf{95.89$\pm$0.62} \\
 & 3 & 66.6$\pm$0.09 & 64.0$\pm$0.21 & 63.7$\pm$0.35 & 62.4$\pm$0.26 & 62.9$\pm$0.26 & 70.0$\pm$0.30 & 65.4$\pm$0.28 & 72.2$\pm$0.26 & 91.4$\pm$0.75 & \textbf{93.22$\pm$1.01} \\
 & 4 & 66.5$\pm$0.10 & 60.9$\pm$0.26 & 63.7$\pm$0.21 & 63.0$\pm$0.23 & 61.7$\pm$0.30 & 63.6$\pm$0.25 & 61.2$\pm$0.21 & 68.8$\pm$0.35 & 90.5$\pm$0.80 & \textbf{92.34$\pm$1.15} \\
 & 5 & 56.3$\pm$0.16 & 56.5$\pm$0.17 & 58.5$\pm$0.27 & 58.2$\pm$0.15 & 57.4$\pm$0.32 & 60.7$\pm$0.30 & 59.3$\pm$0.15 & 60.2$\pm$0.42 & 88.9$\pm$0.80 & \textbf{92.10$\pm$1.69} \\
 & All & 57.6$\pm$0.13 & 57.4$\pm$0.14 & 59.5$\pm$0.23 & 59.0$\pm$0.16 & 58.6$\pm$0.29 & 62.1$\pm$0.36 & 60.4$\pm$0.16 & 61.2$\pm$0.37 & 89.8$\pm$1.05 & \textbf{92.57$\pm$1.53} \\
\hline
\multirow{6}{*}{\makecell{Meeting\\room}}
 & 1 & 76.7$\pm$0.07 & 72.4$\pm$0.15 & 79.5$\pm$0.20 & 75.9$\pm$0.23 & 73.9$\pm$0.30 & 81.6$\pm$0.26 & 76.7$\pm$0.17 & 86.4$\pm$0.21 & 96.8$\pm$0.80 & \textbf{99.62$\pm$0.46} \\
 & 2 & 73.6$\pm$0.10 & 71.2$\pm$0.16 & 68.6$\pm$0.25 & 72.7$\pm$0.18 & 69.8$\pm$0.28 & 71.1$\pm$0.41 & 69.9$\pm$0.24 & 79.9$\pm$0.33 & 92.8$\pm$0.85 & \textbf{95.58$\pm$0.49} \\
 & 3 & 66.8$\pm$0.14 & 60.9$\pm$0.23 & 66.2$\pm$0.20 & 64.3$\pm$0.20 & 64.0$\pm$0.21 & 68.2$\pm$0.34 & 67.5$\pm$0.18 & 72.0$\pm$0.26 & 91.9$\pm$0.75 & \textbf{94.67$\pm$0.25} \\
 & 4 & 66.2$\pm$0.13 & 61.1$\pm$0.21 & 61.9$\pm$0.21 & 59.1$\pm$0.28 & 61.4$\pm$0.26 & 65.5$\pm$0.31 & 62.7$\pm$0.28 & 66.0$\pm$0.38 & 91.2$\pm$0.80 & \textbf{93.96$\pm$0.85} \\
 & 5 & 56.2$\pm$0.24 & 58.0$\pm$0.20 & 58.7$\pm$0.30 & 59.5$\pm$0.15 & 58.8$\pm$0.22 & 63.3$\pm$0.35 & 60.0$\pm$0.18 & 61.2$\pm$0.45 & 89.5$\pm$0.65 & \textbf{93.30$\pm$1.54} \\
 & All & 57.2$\pm$0.21 & 59.2$\pm$0.23 & 59.8$\pm$0.36 & 60.2$\pm$0.14 & 59.6$\pm$0.28 & 64.8$\pm$0.38 & 61.3$\pm$0.15 & 62.1$\pm$0.41 & 90.5$\pm$0.69 & \textbf{93.81$\pm$1.12} \\
\hline
\multicolumn{12}{l}{\footnotesize $^{\dagger}$All: model trained on data from all user counts combined.} \\
\end{tabular}%
}
\end{table*}

We compare the proposed method against nine baseline models spanning four architectural families: (i)~traditional machine learning (ST-RF, a Random Forest with handcrafted statistical features); (ii)~shallow neural networks (MLP, LSTM, CNN-1D, CNN-2D); (iii)~recurrent-attention hybrids (CLSTM, ABLSTM); and (iv)~transformer-based architectures (THAT, ITA\cite{wang2025wifi}).
Table~\ref{tab:comparison} reports the recognition accuracy for all three environments and six user configurations.

Three observations emerge from Table~\ref{tab:comparison}.

\emph{First}, traditional and shallow models (ST-RF, MLP, CNN-1D, CNN-2D) collapse in multi-user scenarios: their accuracy drops from $73$--$81\%$ (1 user) to $55$--$58\%$ (5 users), a $20$--$25$-point decline that reflects their inability to disentangle superimposed CSI patterns from multiple users.

\emph{Second}, recurrent and attention-augmented models (LSTM, CLSTM, ABLSTM, THAT) achieve moderate improvements---CLSTM reaches $63$--$65\%$ with 5 users---but their sequential or global attention mechanisms still operate on the mixed-user signal as a whole, with no mechanism to allocate features to specific user slots.

\emph{Third}, ITA achieves the strongest baseline performance ($89$--$97\%$) by combining InceptionTime multi-scale feature extraction with attention-based recalibration.
However, ITA still trails our method by $3$--$4\%$ in multi-user settings, the gap that UCSA fills: while ITA's attention enhances feature quality, it cannot by itself associate features with individual users, whereas our user-conditioned spatial attention provides data-driven per-user attention maps that directly link features to user slots.

Across all environments, our method achieves the highest accuracy in every user configuration, with the advantage growing as user count increases---from $2$--$3\%$ with 1--2 users to $3$--$4\%$ with 4--5 users---confirming that the synergy of shared spatial attention and user-conditioned attention becomes increasingly valuable as inter-user interference intensifies.

\subsection{Ablation Study}

\begin{table}[!t]
\centering
\caption{Ablation study on attention modules (5 users, meeting room).}
\label{tab:ablation_attention}
\begin{tabular}{lc}
\hline
\textbf{Model variant} & \textbf{Accuracy (\%)} \\
\hline
Full model (SMSA + UCSA + ID Embed.) & \textbf{93.30$\pm$1.54} \\
\quad w/o SMSA              & 85.69$\pm$1.87 \\
\quad w/o UCSA          & 88.07$\pm$2.19 \\
\quad w/o SMSA \& UCSA  & 83.21$\pm$1.82 \\
\hline
\end{tabular}
\end{table}

\begin{table}[!t]
\centering
\caption{Ablation study on user-differentiation strategies (5 users, meeting room).}
\label{tab:ablation_embedding}
\begin{tabular}{lc}
\hline
\textbf{Embedding variant} & \textbf{Accuracy (\%)} \\
\hline
w/ user-ID Embedding     & \textbf{93.30$\pm$1.54} \\
w/o user-ID Embedding    & 88.71$\pm$2.08 \\
w/ Position Embedding      & 94.29$\pm$3.19 \\
\hline
\end{tabular}
\end{table}

We conduct ablation experiments on the 5-user meeting room scenario. Table~\ref{tab:ablation_attention} examines the attention modules, and Table~\ref{tab:ablation_embedding} compares user-differentiation strategies.

Removing SMSA alone causes a $7.6$-point drop ($93.30 \rightarrow 85.69\%$), indicating that shared multi-scale spatial feature enhancement is essential for extracting discriminative CSI representations from the mixed-user signal.
Removing UCSA alone causes a $5.2$-point drop ($93.30 \rightarrow 88.07\%$), demonstrating that user-conditioned spatial attention provides a critical link between shared features and per-user predictions: without it, the model reverts to a single user-agnostic attention map. Removing both modules drops accuracy to $83.21\%$ ($-10.1$ points), indicating that SMSA and UCSA address complementary aspects: SMSA enhances the quality of the shared representation, while UCSA ensures that each user slot attends to the spatial regions most relevant to that user. Together, these results confirm that both modules contribute substantially to recognition accuracy.

Adding user-ID embeddings improves accuracy from $88.71\%$ to $93.30\%$ ($+4.6\%$), indicating that learnable slot-specific identifiers allow the model to associate shared features with individual user slots without any location prior.
Position embedding achieves a marginally higher mean ($94.29\%$) but requires ground-truth location information during inference, which constitutes information leakage and is unavailable in practice.
Moreover, position embedding exhibits substantially higher variance ($\sigma\!=\!3.19$ vs.\ $1.54$), suggesting unstable convergence likely caused by overfitting to specific spatial configurations. Overall, user-ID embedding provides a practical and robust alternative: it matches position embedding's discriminative intent while maintaining stable generalization without requiring prior location knowledge.

Taken together, the ablation results reveal a clear division of labor: SMSA enhances the shared feature representation, while UCSA creates user-specific attention patterns that bridge shared features and per-user predictions. Neither component alone is sufficient, and the dual interaction of user-ID embeddings ensures that user identity information shapes the entire feature extraction pipeline rather than being injected only at the final layer.

\subsection{Attention Map Visualization}

\begin{figure*}[tbp]
    \centering
    \includegraphics[width=\textwidth]{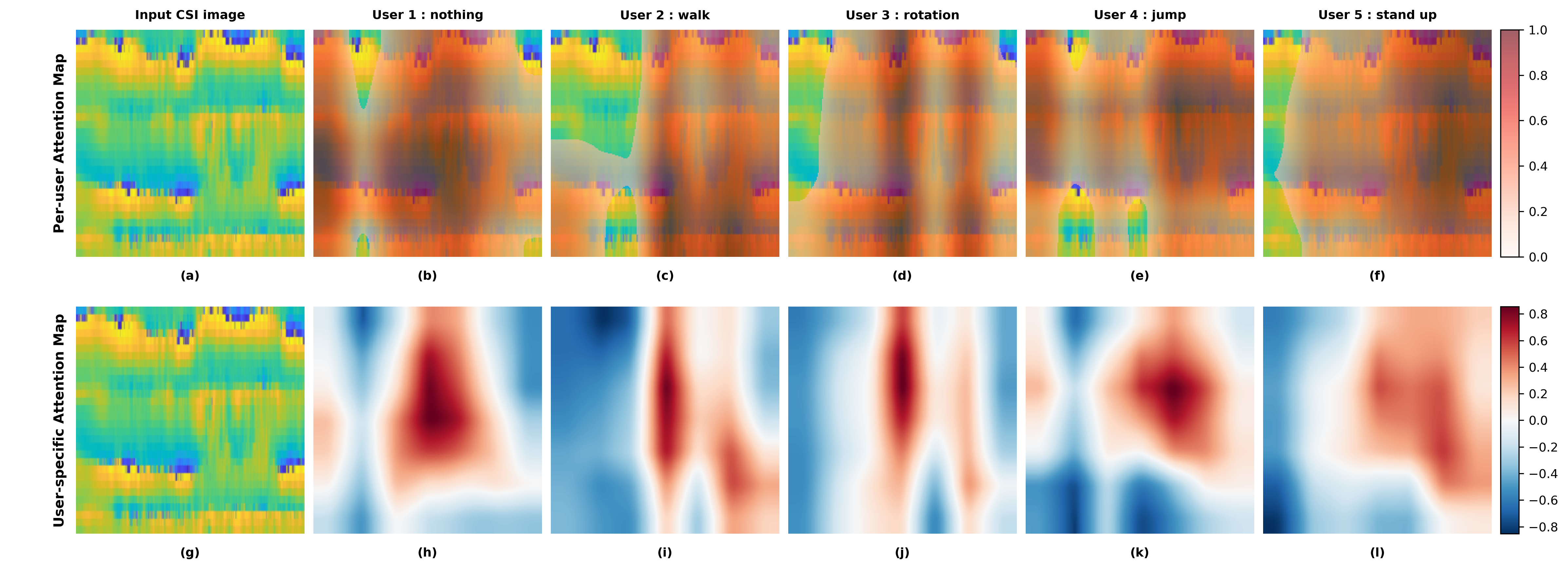}
    \captionsetup{justification=raggedright,singlelinecheck=false}
    \caption{UCSA attention visualization under the 5-user scenario.Top row: per-user attention maps overlaid on the input CSI image. Bottom row: user-specific attention maps $\Delta\mathbf{A}_m $. Red: stronger than base attention; blue: weaker than base attention.}
    \label{fig:attn_overlay}
\end{figure*}

To illustrate how UCSA produces user-conditioned spatial attention, we visualize the per-user attention maps under the 5-user meeting room scenario.
Fig.~\ref{fig:attn_overlay} presents two complementary views of the same attention data.

The top row shows each user's absolute attention map overlaid on the input CSI image, answering the question: ``Which spatial regions does this user attend to?''.
The attention maps exhibit clear spatial structure, concentrating on physically meaningful regions of the CSI image rather than being randomly distributed, which confirms that UCSA's spatial attention captures motion-relevant patterns.

The \emph{bottom row} shows the \emph{difference} between each per-user attention map and the shared base attention, answering the question: ``Which spatial regions does this user attend to \emph{differently} from the common baseline?''.
Formally, the user-specific map for user $m$ is:
\begin{equation}
\Delta\mathbf{A}_m = \mathbf{A}_m^{\text{UCSA}} - \mathbf{A}^{\text{base}},
\end{equation}
where $\mathbf{A}_m^{\text{UCSA}} = \sigma(\boldsymbol{\gamma}_m \odot \mathbf{A}^{\text{base}} + \boldsymbol{\beta}_m)$ is the FiLM-modulated attention for user $i$ and $\mathbf{A}^{\text{base}}$ is the shared base attention.
Red regions indicate where the user attends more strongly than the baseline, and blue regions indicate where the user attends less strongly.

The user-specific maps in the bottom row reveal distinct spatial patterns across the five user slots: P1 (nothing) concentrates on the central region, P2 (walk) exhibits a vertical stripe pattern in the left portion, P4 (jump) shows a concentrated focal region in the center-right area, and P5 (stand up) highlights the right portion.
These differences are not attributable to the input alone (all users share the same CSI image) but arise from the FiLM modulation of the user-ID embeddings, which shifts each user's attention relative to the common baseline.
Notably, regions that all users attend to similarly (e.g., the center of the image, where the strongest motion signature is located) appear white in the difference map, while the colored regions reveal the user-specific adjustments introduced by UCSA.
This visualization  provides  direct  evidence  that UCSA  achieves  genuine user conditioning: the same input produces different spatial attention patterns for different user slots, and these differences are consistent with the physical expectation that users performing different activities generate distinct spatial signatures in the CSI image.
%This visualization illustrates that UCSA produces different spatial attention patterns for different user slots. However, since the users also perform different activities, the observed differences cannot be attributed solely to user identity. Nevertheless, combined with the user-slot confusion matrix and the ablation results, these visualizations support the effectiveness of UCSA in generating slot-specific attention.

\section{CONCLUSION}

In this work, we have designed a WiFi-based multi-user activity recognition framework that effectively addresses the coupled challenges of spatial feature quality and user differentiation under concurrent-user interference. The system first employs a CSI-ratio preprocessing scheme to eliminate common phase offsets across receive antennas, converting raw CSI into differential phase-ratio images. The network then incorporates SMSA module for shared multi-scale spatial feature enhancement and UCSA module that generates per-user attention maps by conditioning on learnable user-ID embeddings via FiLM-style modulation, creating a deep coupling between spatial feature enhancement and user differentiation. Experiments on the WiMANS benchmark across three indoor environments with up to five concurrent users demonstrate that the proposed method outperforms nine baselines in all settings, achieving an average accuracy of over 93\% across three environments even with five users. Ablation studies further confirm that both SMSA and UCSA contribute substantially and complementarily: SMSA enhances the shared feature representation, while UCSA bridges shared features and per-user predictions through user-aware spatial attention. The current evaluation is limited to a single benchmark with five participants and predefined locations; extending the framework to larger-scale, more diverse environments with dynamic user arrivals remains for future work.

\bibliographystyle{IEEEtran}
\bibliography{myre.bib}
% \end{multicols}
\end{document}